\documentclass[graybox]{svmult}

\usepackage{type1cm}        
\usepackage{makeidx}         
\usepackage{graphicx}        
\usepackage{multicol}        
\usepackage[bottom]{footmisc}

\usepackage{newtxtext}       %
\usepackage[varvw]{newtxmath}       

\newcommand{\rf}[1]{(\ref{#1})}
\usepackage{multirow}
\usepackage{makecell}

\newcolumntype{C}[1]{>{\centering\let\newline\\\arraybackslash\hspace{0pt}}m{#1}}

\makeindex             

\begin{document}

\title*{Dynamical dark energy with AdS-dS transitions vs. Baryon Acoustic Oscillations at $z = 2.3 - 2.4$}
\titlerunning{Dynamical dark energy with AdS-dS transitions}
\author{\"{O}zg\"{u}r Akarsu \\ Maxim Eingorn \\ Leandros Perivolaropoulos \\ A. Emrah Y\"{u}kselci \\ Alexander Zhuk}
\authorrunning{Akarsu, Eingorn, Perivolaropoulos, Y\"{u}kselci, Zhuk}
\institute{\"{O}zg\"{u}r Akarsu \at Department of Physics, Istanbul Technical University, Maslak 34469 Istanbul, T\"{u}rkiye \email{akarsuo@itu.edu.tr}
\and Maxim Eingorn \at Department of Mathematics and Physics, North Carolina Central University, \\1801 Fayetteville St., Durham, North Carolina 27707, U.S.A. \email{maxim.eingorn@gmail.com} \and Leandros Perivolaropoulos \at Department of Physics, University of Ioannina, GR-45110, Ioannina, Greece \email{leandros@uoi.gr} \and A. Emrah Y\"{u}kselci \at Department of Physics, Istanbul Technical University, Maslak 34469 Istanbul,  T\"{u}rkiye \email{yukselcia@itu.edu.tr} \and Alexander Zhuk \at Center for Advanced Systems Understanding, Untermarkt 20, 02826 G\"{o}rlitz, Germany; Helmholtz-Zentrum Dresden-Rossendorf, Bautzner Landstra\ss e 400, 01328 Dresden, Germany; Astronomical Observatory, Odesa I.I. Mechnikov National University, Dvoryanskaya St. 2, Odesa 65082, Ukraine \email{ai.zhuk2@gmail.com}}
%
%
\maketitle

\abstract{
In this paper, written in memory of \textit{Alexei Starobinsky}, we discuss the observational viability of the Ph-$\Lambda_{\rm s}$CDM model --- a dynamical dark energy scenario based on a phantom scalar field undergoing an anti-de Sitter (AdS) to de Sitter (dS) transition --- and revisit the Sahni–Shtanov braneworld model in light of updated BAO Ly-$\alpha$ data at $z \sim 2.3$. Both models are able to remain consistent with Planck CMB data while offering potential resolutions to the $H_0$ tension. In both cases, the expansion rate $H(z)$ is suppressed relative to Planck-$\Lambda$CDM at high redshift and enhanced at low redshift, while remaining consistent with the comoving distance to recombination as estimated by Planck-$\Lambda$CDM. Comparing model predictions with BAO-inferred values of $H(z)$, we find that SDSS Ly-$\alpha$ data at $z \approx 2.33$ mildly favor such dynamical models, whereas the recent DESI Ly-$\alpha$ measurements agree more closely with $\Lambda$CDM. Although current high-redshift BAO data do not decisively favor one model over another, our findings illustrate how frameworks originally developed to address earlier anomalies --- such as the braneworld scenario --- may gain renewed relevance in confronting today's cosmological tensions.
}

\section{Introduction}
\label{sec:intro}

This paper, \textit{dedicated to the memory of Alexei Starobinsky}, is inspired by the article~\cite{Starob}, coauthored by Varun Sahni, Arman Shafieloo, and Alexei Starobinsky himself. In that work, the authors investigated the dynamics of the Universe within the braneworld model framework~\cite{Brane}, and observed that the Hubble parameter in this model takes smaller values than in the standard $\Lambda$CDM model over an extended period in the past. A similar behavior emerged in our recent study~\cite{our}, where we introduced the Ph-$\Lambda_{\rm s}$CDM model, in which dark energy is modeled by a phantom scalar field with a hyperbolic tangent potential. Owing to the specific form of this potential and the phantom nature of the scalar field, our framework naturally realizes bottom-up AdS-to-dS transitions. The model is derived from first principles — grounded in a well-defined action — and offers a physical realization of the phenomenological $\Lambda_{\rm s}$CDM scenario, wherein the effective cosmological constant transitions from negative to positive values as conjectured based on findings in graduated dark energy (gDE)~\cite{gDE,LsCDM1,LsCDM2,LsCDM3}. We demonstrate that, on the one hand, this approach maintains consistency with the full Planck 2018 CMB data~\cite{Planck}, and on the other, yields a present-day Hubble parameter $H_0$ in agreement with SH0ES measurements~\cite{SH0ES}, thereby offering a potential resolution to the longstanding $H_0$ tension in cosmology.

We also show that the equation-of-state (EoS) parameter $\omega_\phi$ of the phantom field evolves from $\omega_\phi \gtrsim -1$ before the transition — when its energy density is negative ($\varepsilon_\phi < 0$) — to $\omega_\phi \lesssim -1$ after the transition, when the energy density becomes positive. Notably, $\omega_\phi$ exhibits a pole (a so-called safe singularity) at the transition point, where the total energy density of the scalar field vanishes. An analogous evolution of the effective EoS parameter is found in the braneworld model examined in Ref.~\cite{Starob}, which is based on the framework introduced in Ref.~\cite{Brane}. Moreover, at redshift $z = 2.34$, this braneworld model provides a better fit to the line-of-sight baryon acoustic oscillation (BAO) signal from Ly-$\alpha$ forest data in SDSS DR11~\cite{Delubac et al 2015} than the standard $\Lambda$CDM model~\cite{Starob}. According to Refs.~\cite{Delubac et al 2015,Aubourg et al 2015}, BAO measurements at this redshift — derived from Ly-$\alpha$ forest auto-correlation and Ly-$\alpha$--quasar cross-correlation — indicate a lower value of the Hubble parameter $H(z)$ than predicted by the standard $\Lambda$CDM model. This deviation suggests a possible preference for cosmological scenarios in which the expansion rate at high redshift is suppressed relative to $\Lambda$CDM. This is precisely the behavior exhibited by the braneworld model in Ref.~\cite{Starob}. In our dynamical dark energy model~\cite{our}, we similarly find that $H(z)$ is lower than in $\Lambda$CDM at sufficiently high redshifts (approximately $z \gtrsim 1$). This naturally raises the question of whether our model provides a better fit to the BAO observations in this redshift regime. The present work is dedicated to addressing this question. As we will show, certain BAO measurements — particularly those based on SDSS Ly-$\alpha$ data — exhibit a preference for such behavior, while the most recent DESI results remain more consistent with the $\Lambda$CDM prediction.

The paper is organized as follows. In Sec.~2, we present a brief description of dynamical dark energy in the Ph-$\Lambda_{\rm s}$CDM model, as well as in the generalized Sahni--Shtanov braneworld model. There, we perform numerical computations of the Hubble parameter $H(z)$ such that both models are consistent with the CMB data on the one hand, and satisfy the SH0ES determination of $H_0$ on the other. In Sec.~3, we compile recent BAO Ly-$\alpha$ measurements and compare the corresponding values of the Hubble parameter with those predicted by the Ph-$\Lambda_{\rm s}$CDM model, the braneworld model, and the standard $\Lambda$CDM model. Finally, in Sec.~4, we summarize and discuss our findings.

\section{Dynamical dark energy with AdS-dS transitions}
\label{sec:2}

In this section, we provide a brief description of the dynamical dark energy model; a detailed analysis can be found in Ref.~\cite{our}. The matter content of the Universe is taken to consist of radiation, pressureless matter --- i.e., cold dark matter (CDM) and baryons --- and a minimally coupled scalar field $\phi$ with a potential $V(\phi)$, described by the action
\begin{equation}
	S_{\phi} = \int {\rm d}^4x\,\sqrt{-g} \left[ \frac{\xi}{2} g^{ik} \partial_{i} \phi \partial_{k} \phi - V(\phi) \right] \, .
\end{equation}
We consider this model at the background level, where the scalar field depends only on time, and the background metric is taken to be spatially flat Friedmann--Lema\^{\i}tre--Robertson--Walker (FLRW). The parameter $\xi = +1$ corresponds to quintessence (a canonical scalar field with positive kinetic energy), while $\xi = -1$ corresponds to a phantom field (a scalar field with negative kinetic energy). In what follows, we focus exclusively on the phantom case, i.e., $\xi = -1$.

The scalar field potential is taken to be
\begin{equation}
	V(\phi) = \dfrac{\Lambda(\xi_1 + 1)}{2} - \dfrac{\Lambda(\xi_1 - 1)}{2} \tanh\left[\sqrt{\kappa}\, \nu \left(\phi - \phi_c\right)\right],
	\label{eqn:gen_pot}
\end{equation}
where $\Lambda > 0$, $\nu$ is a rapidity parameter characterizing the transition rate, and $\kappa \equiv 8\pi G_N / c^4$, with $G_N$ being Newton’s gravitational constant and $c$ the speed of light. The parameter $\phi_c$ denotes the inflection point of the potential, marking the midpoint of the transition where the hyperbolic tangent changes sign. This potential exhibits the following asymptotic behavior: $V(\phi \to -\infty) \to \Lambda \xi_1$ and $V(\phi \to +\infty) \to \Lambda$. Therefore, for $\xi_1 < 0$, the phantom field — with its negative kinetic energy — evolves from a region with negative potential energy to one with positive potential energy, thereby realizing an AdS-to-dS transition. The case $\xi_1 = -1$ describes a mirror transition from $-\Lambda$ to $+\Lambda$, providing a natural dynamical realization of the $\Lambda_{\rm s}$CDM scenario~\cite{gDE,LsCDM1,LsCDM2,LsCDM3}. In contrast, the case $\xi = 0$, $\xi_1 = 1$ recovers the standard $\Lambda$CDM model.

The full system of equations of motion for the phantom field $\phi(t)$ and the scale factor $a(t)$ is presented in Ref.~\cite{our}. We solve these equations numerically with initial conditions $\phi_{\rm in} = \dot{\phi}_{\rm in} = 0$, where $\phi_{\rm in} < \phi_{\rm c}$. In addition, we impose the following constraints on the present-day values of the physical densities of matter (CDM + baryons):
\begin{equation}
	\Omega_{\rm m0} h^2 = 0.14314,
	\label{eq:Omega_M0}
\end{equation}
and radiation:
\begin{equation}
	\Omega_{\rm r0} h^2 = 2.469 \times 10^{-5} \times \left[1 + \frac{7}{8} \left(\frac{4}{11}\right)^{\!4/3} N_{\rm eff} \right],
	\label{eq:Omega_r0}
\end{equation}
where $h \equiv H_0 / (100~{\rm km\,s}^{-1}\,{\rm Mpc}^{-1})$ is the dimensionless reduced Hubble constant, and $N_{\rm eff} = 3.046$, in accordance with the standard model of particle physics. The value of the physical matter density in Eq.~\eqref{eq:Omega_M0} is chosen to match the Planck-CMB estimates~\cite{Planck}. Furthermore, the comoving angular diameter distance to the surface of last scattering, $D_M(z_*)$, with $z_* \approx 1090$, is tightly constrained by the Planck CMB power spectra within the $\Lambda$CDM framework~\cite{D_M}:
\begin{align}
	D_M(z_*) = \int_0^{z_*} \frac{c}{H(z)}\,{\rm d}z = 13869.57~\mathrm{Mpc},
	\label{eq:D_M}
\end{align}
which ensures that the sound horizon at recombination, $r_*$, remains the same as in the Planck-$\Lambda$CDM model. This is because the angular scale of the sound horizon, $\theta_* = r_*/D_M(z_*)$, is measured with high precision and is nearly model-independent. Thus, the constraints given in Eqs.~\eqref{eq:Omega_M0}--\eqref{eq:D_M} ensure consistency with the CMB power spectra~\cite{Planck}. For each solution $H(z)$, we fix the present-day Hubble parameter to $H_0 = 73.04~{\rm km\,s^{-1}\,Mpc^{-1}}$, in accordance with the SH0ES measurement~\cite{SH0ES}, and verify that it satisfies Eq.~\eqref{eq:D_M}. This procedure uniquely determines the value of $\phi_{\rm c}$ and the redshift of the AdS--dS transition, $z_{\rm t}$, for each solution.

\begin{figure}[t]
	\includegraphics[width=1\linewidth]{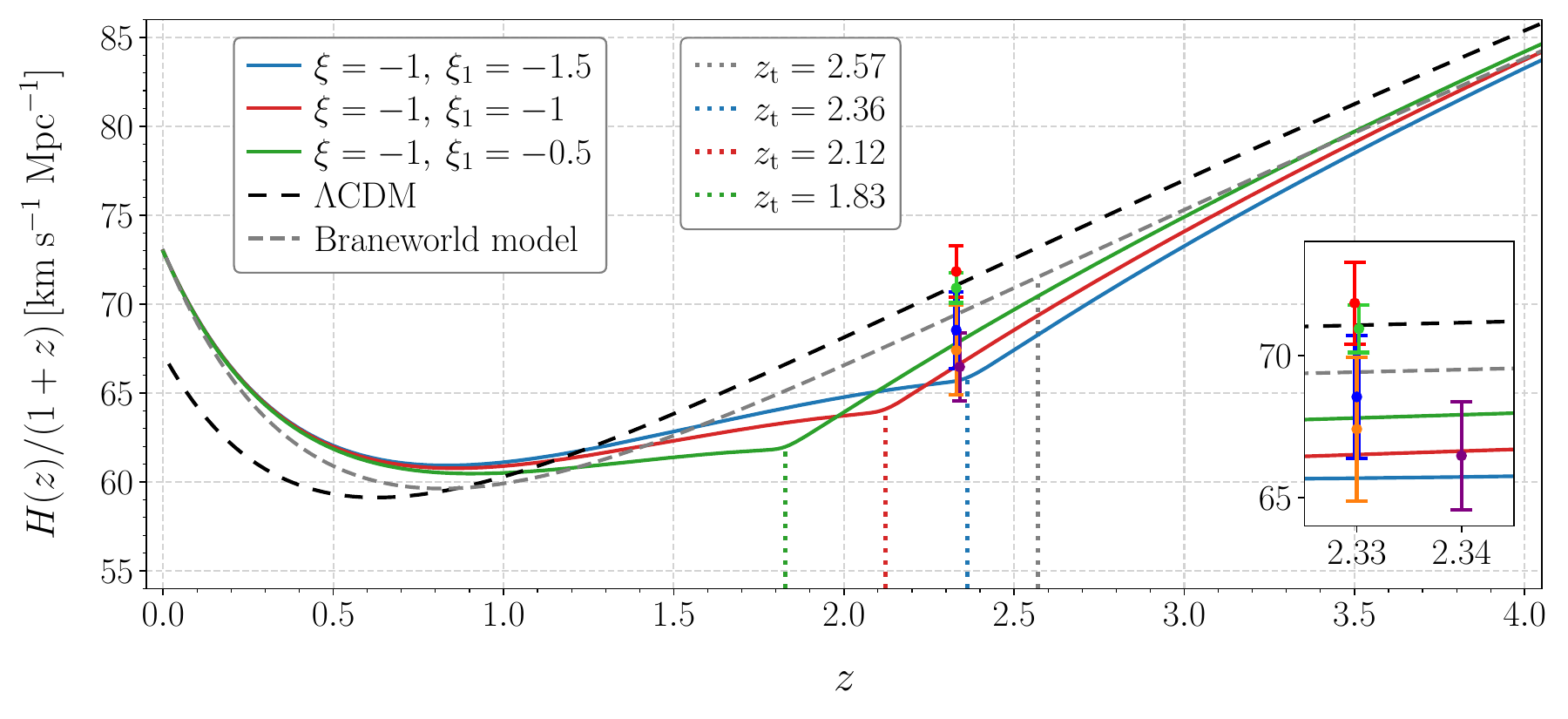}
	\vspace{-4mm}
	\caption{The evolution of the co-moving Hubble parameter, $\dot a=H(z)/(1+z)$, for 
		selected models. All curves are continuous and smooth. The vertical dotted lines determine the transition redshifts $z_{\rm t}$. They correspond to the inflection points of potential $V(\phi)$ in the case of the phantom model and the point where the density of dark energy $\varepsilon_{\rm{de}}$ changes it sign in the case of the braneworld model. The phantom cases, i.e. $\xi=-1$, and the braneworld model satisfy the present-day value for the Hubble parameter in accordance with the SH0ES, that is $73.04~{\rm km\, s^{-1}\, Mpc^{-1}}$. The black dashed line corresponds to $\Lambda$CDM solution ($\xi=0, \xi_1=1$) which gives $H_0=67.22~{\rm km\, s^{-1}\, Mpc^{-1}}$. We also indicate 5 selected values of $H(z)/(1+z)$ with error bars corresponding to BAO measurements. The  purple one was obtained from SDSS-III (DR11) for $z=2.34$~\cite{Delubac et al 2015}. Four others correspond to $z=2.33$. The orange and blue were get from SDSS-IV for two different types of measurements~\cite{Alam et al 2021} and the bright red and green were obtained from DESI BAO measurements~\cite{Adame et al 2024} and~\cite{Karim et al 2025}, respectively.
        }
	\label{fig:H_over_z} 
\end{figure}

Figure~\ref{fig:H_over_z} shows the result of the numerical solution for the comoving Hubble parameter, ${\rm d}a/{\rm d}t = H(z)/(1+z)$, for three phantom models with $\xi_1 = -1.5, -1, -0.5$ (colored curves), as well as for the standard $\Lambda$CDM model (black dashed curve). The rapidity parameter is fixed at $\nu = 100$. All phantom models are calibrated to yield $H_0 = 73.04~{\rm km\,s^{-1}\,Mpc^{-1}}$, in accordance with the SH0ES measurement~\cite{SH0ES}. In contrast, the standard $\Lambda$CDM model is shown with $H_0 = 67.22~{\rm km\,s^{-1}\,Mpc^{-1}}$, which represents the upper limit compatible with the constraints in Eqs.~\eqref{eq:Omega_M0},~\eqref{eq:Omega_r0}, and~\eqref{eq:D_M} under the $\Lambda$CDM framework.

All phantom model curves are continuous and smooth, with clearly defined transition redshifts $z_{\rm t}$, which correspond to the inflection points of the scalar field potential given in Eq.~\eqref{eqn:gen_pot}. The transition redshifts are $z_{\rm t} = 2.36,\, 2.12,\, 1.83$ for $\xi_1 = -1.5,\, -1,\, -0.5$, respectively. These redshifts mark characteristic points in the evolution of the model parameters during the AdS--dS transition, as illustrated in Fig.~\ref{fig:H_over_z}. It is worth noting that another special point in the model is the redshift $z_\dagger$, defined by the condition that the energy density of the scalar field vanishes:
\[
\varepsilon_{\phi}(z = z_\dagger) = -\frac{\dot{\phi}^2}{2 c^2} + V(\phi) = 0,
\]
where the overdot denotes a derivative with respect to cosmic time. The EoS parameter $\omega_{\phi}$ exhibits a pole at $z_\dagger$. Numerical calculations show that $z_\dagger < z_{\rm t}$, which is explained by the influence of the kinetic term.

Figure~\ref{fig:H_over_z} demonstrates that the Hubble parameter in the considered phantom models lies below that of the $\Lambda$CDM model for $z \gtrsim 1.1$--$1.2$. However, in order to satisfy the fixed value of the integral in Eq.~\eqref{eq:D_M}, the phantom model curves must lie above the $\Lambda$CDM curve for $z \lesssim 1.1$. This behavior explains why the phantom field models yield $H_0$ values consistent with the SH0ES measurement.

In Fig.~\ref{fig:H_over_z}, we also include the curve corresponding to the braneworld model (gray dashed line). The Hubble parameter in a spatially flat brane cosmology satisfies the equation~\cite{Starob,Brane}:
\begin{equation}
\frac{H^2(z)}{H_0^2} = \Omega_{\rm m0}(1+z)^3 + \Omega_{\rm r0}(1+z)^4 +
\tilde\Omega_{\rm de}(z),
\label{Hubble brane} 
\end{equation}
where $\Omega_{\rm m0} = \varepsilon_{\rm m0} / \varepsilon_{\rm cr0}$, $\Omega_{\rm r0} = \varepsilon_{\rm r0} / \varepsilon_{\rm cr0}$ (we additionally turned on radiation), and $\varepsilon_{\rm cr0} = 3H_0^2 / (\kappa c^2)$. The effective dark energy term is given by
\begin{equation}
\tilde\Omega_{\rm de}(z) \equiv \frac{\varepsilon_{\rm de}}{\varepsilon_{\rm cr0}} = \Omega_\Lambda + 2\Omega_l - 2\sqrt{\Omega_l} \sqrt{ \Omega_{\rm m0}(1+z)^3 + \Omega_{\rm r0}(1+z)^4 + \Omega_\Lambda + \Omega_l },
\label{DE brane} 
\end{equation}
where $\Omega_\Lambda = \Lambda / \varepsilon_{\rm cr0}$, and $\Lambda$ is the effective brane tension associated with a three-dimensional brane embedded in a 4+1-dimensional bulk spacetime. The cosmological parameter $\Omega_l = 1/(l_c^2 H_0^2)$ is defined in terms of a characteristic length scale $l_c$, which is given by the ratio $l_c = M_{\rm Pl(4)}^2 /  M_{\rm Pl(5)}^3$. Here, $M_{\rm Pl(4)}$ and $M_{\rm Pl(5)}$ denote the four-dimensional and five-dimensional reduced Planck masses, respectively. The scale $l_c$ characterizes the transition between four-dimensional and higher-dimensional gravitational behavior in the braneworld scenario. It is convenient to use the following consistency relation for the present-day density parameters~\cite{Brane}:
\begin{equation}
\Omega_\Lambda = 1 - \Omega_{\rm m0} - \Omega_{\rm r0} + 2\sqrt{\Omega_l}.
\label{brane constr} 
\end{equation}

To integrate Eq.~\eqref{Hubble brane} under the constraint~\eqref{brane constr}, we impose the same conditions as in Eqs.~\eqref{eq:Omega_M0}--\eqref{eq:D_M}, and adopt the SH0ES value $H_0 = 73.04~{\rm km\,s^{-1}\,Mpc^{-1}}$~\cite{SH0ES}. These additional constraints (which were not employed in the original analysis of Ref.~\cite{Starob}), when combined with Eq.~\eqref{brane constr}, fully determine the model parameters. For example, fixing $H_0 = 73.04~{\rm km\,s^{-1}\,Mpc^{-1}}$ in accordance with the SH0ES mean value yields $\Omega_l \approx 0.023$. In turn, choosing $\Omega_l = 0.025$, as done in Ref.~\cite{Starob}, leads to $H_0 = 73.24~{\rm km\,s^{-1}\,Mpc^{-1}}$, which remains within the SH0ES error bars. As in the original work~\cite{Starob}, we find that at high redshifts the Hubble parameter in the braneworld model lies significantly below that of the standard $\Lambda$CDM scenario. This behavior originates from the negative contribution of the dark energy density $\varepsilon_{\rm de}(z)$ in Eq.~\eqref{Hubble brane}, which reduces the total energy density driving the expansion for $z > z_{\rm t}$. At the critical redshift $z_{\rm t} = 2.57$, $\varepsilon_{\rm de}$ vanishes and subsequently becomes positive, reversing its effect on the expansion rate. To satisfy the CMB constraint on the comoving angular diameter distance, Eq.~\eqref{eq:D_M}, the model must compensate for this early-time suppression of $H(z)$ by producing an enhanced expansion rate at lower redshifts. As a result, the braneworld Hubble curve eventually overtakes the $\Lambda$CDM curve, reaching a present-day value of $H_0$ consistent with the SH0ES measurement. This transition was not captured in Ref.~\cite{Starob}, as the condition~\eqref{eq:D_M} was not imposed in that analysis. In summary, much like our phantom field scenario, the inclusion of CMB-integrated distance constraints enables the braneworld model to reconcile the high-redshift suppression of the expansion rate with the high local value of $H_0$, thereby maintaining consistency with both Planck and SH0ES observations.

To conclude this section, we emphasize several important properties of our dynamical dark energy model. It is well known that phantom fields violate the null energy condition (NEC), since their inertial mass density — defined as $\varrho_{\phi} \equiv \varepsilon_{\phi} + p_{\phi}$ — is negative. Moreover, the energy density $\varepsilon_{\phi}$ itself can also become negative, particularly when the scalar field potential is negative. However, in the presence of additional matter components such as cold dark matter and radiation, the total inertial mass density and total energy density can remain non-negative at all times. Our calculations~\cite{our} confirm that this is indeed the case in our model. Therefore, the weak energy condition (and consequently the NEC) is not violated in the combined system. Another common concern with phantom models is the emergence of a big rip singularity: in scenarios where the phantom field has positive energy density, the expansion typically leads to a divergence of the scale factor and energy density within finite cosmic time~\cite{bigrip0,bigrip1,bigrip1.1,bigrip2,bigrip3}. However, if the late-time geometry asymptotes to de Sitter spacetime, the big rip can be avoided~\cite{bigrip4}. It is also crucial to recognize that the negative kinetic term intrinsic to phantom fields generically leads to dynamical instabilities, unless the scalar field potential is appropriately bounded from above. Thus, both the big rip singularity and the background instability associated with negative kinetic energy can be mitigated through a carefully chosen potential. The potential given in Eq.~\eqref{eqn:gen_pot} is specifically constructed to address these issues, ensuring both the stability of the model and the avoidance of singular behavior. A further potential concern is the instability of perturbations in phantom field models, particularly if the second derivative of the potential, $ {\rm d}^2 V / {\rm d} \phi^2 $, becomes positive (convex)~\cite{CHT}. In our case, however, the potential remains convex only for a very limited duration~\cite{our}, during which time any unstable modes do not grow appreciably.


\section{Baryon acoustic oscillations at $z=2.3-2.4$}
\label{sec:3}

BAO observations allow us to estimate the ratio $D_{\rm H}/r_{\rm d}$, where $D_{\rm H}(z) = c/H(z)$ is the Hubble distance at redshift $z$, and $r_{\rm d}$ is the comoving sound horizon at the drag epoch, approximately $z \approx 1060$~\cite{Planck}. According to the Planck-$\Lambda$CDM results (see Table 1 in~\cite{Planck}), the best-fit value is $r_{\rm d} = 147.049 \approx 147.05$ Mpc. Given a measurement of the ratio $\frac{D_{\rm H}}{r_{\rm d}} \equiv A(z)$ at redshift $z$, and a known value of $r_{\rm d}$, the Hubble parameter at that redshift can be inferred as:
\begin{equation}
H(z) = \frac{1}{A(z)} \frac{147.05~\mathrm{Mpc}}{r_{\rm d}} \times 2038.71~\mathrm{km\,s^{-1}\,Mpc^{-1}},
\label{Hubble}
\end{equation}
where the prefactor arises from converting between distance and Hubble units, scaled to the reference value $r_{\rm d} = 147.05~\mathrm{Mpc}$.

In Table~\ref{tab:Table1}, we compile measurements of the ratio $D_{\rm H}/r_{\rm d}$ from BAO observations at redshifts $z = 2.3$–$2.4$, focusing specifically on Ly-$\alpha$ forest–based analyses~\cite{Delubac et al 2015,Bautista et al 2017,Bourboux et al 2017,Bourboux et al 2020,Alam et al 2021,Adame et al 2024,Karim et al 2025} (see also~\cite{Aubourg et al 2015,Sapone and Nesseris}). BAO measurements at these redshifts have had a significant impact on cosmology, as they provide a powerful and independent test of the flat $\Lambda$CDM model inferred from Planck CMB data~\cite{Bourboux et al 2020}. The values of $D_{\rm H}/r_{\rm d}$ were extracted from three types of analyses: (i) Ly-$\alpha$ forest auto-correlation (Ly-$\alpha$--Ly-$\alpha$), (ii) cross-correlation between quasars and the Ly-$\alpha$ forest (Ly-$\alpha$--QSO), and (iii) a combined analysis incorporating both auto- and cross-correlations.

As shown in Table~\ref{tab:Table1}, measurements of $D_{\rm H}/r_{\rm d}$ from different SDSS data releases (DR) exhibit slight variations. Even within the same data release, the results depend on the specific analysis method—auto-correlation, cross-correlation, or their combination. A more noticeable difference arises when comparing the SDSS results with those from the recent DESI data releases.\footnote{We include the value $D_{\rm H}/r_{\rm d}(z = 2.330) = 8.632 \pm 0.098 \pm 0.026$ reported in Ref.~\cite{Karim et al 2025} from DESI DR2, where the first and second uncertainties correspond to statistical and theoretical systematic errors, respectively. In our analysis, we combine these in quadrature to obtain the total uncertainty.} The corresponding values of the Hubble parameter $H(z)$ are calculated using Eq.~\eqref{Hubble}, adopting the Planck-$\Lambda$CDM best-fit value $r_{\rm d} = 147.09 \pm 0.26~\mathrm{Mpc}$ (from TT, TE, EE + lowE + lensing)~\cite{Planck}. In Table~\ref{fig:Table2}, we also report the theoretical predictions for $H(z)$ at redshifts $z = 2.33$, $2.34$, and $2.40$, computed for the Ph-$\Lambda_{\rm s}$CDM model (with several choices of $\xi_1$), the braneworld model, and the standard $\Lambda$CDM model.

\begin{table}[t]
    \caption{We present the values of $D_{\rm H}/r_{\rm d}$ obtained by different methods of BAO measurements for selected numbers of redshifts $z$. The corresponding $H(z)$ values, the last column, are calculated via equation \rf{Hubble} using the Planck 2018 value of $r_{\rm d}=147.09\pm0.26$ Mpc (TT,TE,EE+lowE+lensing).}
	\label{tab:Table1}
	\begin{tabular}{p{3.5cm}p{2cm}p{3.5cm}p{2.3cm}}
		\hline
		& & &  \\[-2mm]
		Reference & Survey & $D_H/r_d(z)$  & $H(z)$  km/s/Mpc   \\[1mm] 
		\hline
		& & & \\[-1mm] 
		\makecell[l]{Delubac et al. 2015, \\ A\&A 574 (2015) A59} & \makecell[l]{SDSS-III,\\DR 11} & \makecell[l]{$Ly\alpha-Ly\alpha,\;z=2.34$\\$9.18\pm0.28$} & \makecell[l]{$222.02\pm6.78$\\$z=2.34$} \\[4mm]
		
		\makecell[l]{Bautista et al. 2017,\\A\&A 603(2017) A12} & \makecell[l]{SDSS-III,\\DR 12} & \makecell[l]{$Ly\alpha-Ly\alpha,\;z=2.33$\\$9.07\pm0.31$}  & \makecell[l]{$224.71\pm7.69$\\$z=2.33$} \\[4mm]
		
		\multirow{3}{*}{\makecell[l]{Bourboux et al. 2017,\\A\&A 608(2017) A130}} & \multirow{3}{*}{\makecell[l]{SDSS-III,\\DR 12}} & \makecell[l]{$Ly\alpha-Ly\alpha,\;z=2.4$\\$8.94\pm0.22$} & \makecell[l]{$227.98\pm5.62$\\$z=2.4$} \\[4mm]
		& & \makecell[l]{$Ly\alpha-QSO,\;z=2.4$\\$9.01\pm0.36$} & \makecell[l]{$226.21\pm9.05$\\$z=2.4$} \\[4mm]
		
		\multirow{6}{*}{\makecell[l]{Alam et al. 2021,\\PRD 103(2021)083533; \\ Bourboux et al. 2020, \\ ApJ 901 (2020) 153   
		}} & \multirow{6}{*}{\makecell[l]{SDSS-IV}} & \makecell[l]{$Ly\alpha-Ly\alpha,\;z=2.33$\\$8.93\pm0.28$} & \makecell[l]{$228.24\pm7.17$\\$z=2.33$} \\[4mm]
		& & \makecell[l]{$Ly\alpha-QSO,\;z=2.33$\\$9.08\pm0.34$} & \makecell[l]{$224.47\pm8.41$\\$z=2.33$} \\[4mm]
		& & \makecell[l]{$Ly\alpha-Ly\alpha+$\\$Ly\alpha-QSO,\;z=2.33$\\$8.99\pm0.19$} & \makecell[l]{$226.71\pm4.81$\\$z=2.33$} \\[4mm]
		
		\makecell[l]{Adame et al. 2024,\\JCAP 02 (2025) 021} & \makecell[l]{DESI DR1 BAO} & \makecell[l]{$Ly\alpha-QSO,\;z=2.33$\\$8.52\pm0.17$} & \makecell[l]{$239.22\pm4.79$\\$z=2.33$} \\[4mm]

        \makecell[l]{Karim et al 2025,\\arXiv: 2503.14739} & \makecell[l]{DESI DR2 BAO} & \makecell[l]{$Ly\alpha-Ly\alpha+$\\$Ly\alpha-QSO,\;z=2.33$\\$8.632\pm0.101$} & \makecell[l]{$236.12\pm2.79$\\$z=2.33$} \\[4mm]
		\hline
	\end{tabular}
\end{table}

Based on the observational values summarized in Table~\ref{tab:Table1} and their comparison with theoretical predictions in Table~\ref{fig:Table2}, we may ask: which of the models—the standard $\Lambda$CDM, our Ph-$\Lambda_{\rm s}$CDM model, or the braneworld scenario—provides a better fit to the high-redshift BAO data? To address this qualitatively, we have plotted in Fig.~\ref{fig:H_over_z} the BAO-inferred Hubble parameter values from three major surveys: SDSS-III DR11 (used in~\cite{Starob}), SDSS-IV, and DESI. A visual comparison suggests that the SDSS data tend to favor models in which dark energy transitions from negative to positive values during cosmic evolution, such as the Ph-$\Lambda_{\rm s}$CDM and braneworld models. In contrast, the most recent DESI measurement at $z \sim 2.33$ shows better agreement with the standard $\Lambda$CDM prediction. Thus, while the current data do not yield a definitive preference for one model over another, they highlight interesting differences in model performance across surveys. Of course, a rigorous conclusion requires a full statistical analysis --- such as a likelihood-based parameter estimation using the full BAO datasets and covariance matrices, along with the CMB data --- which lies beyond the scope of this brief paper.

\begin{table}[t]
    \caption{The values 
    of the Hubble parameter $H(z)$ (in $\rm{km}\, \rm{sec}^{-1} \, \rm{Mpc}^{-1}$) calculated for Ph-$\Lambda_{\rm s}$CDM model in three different cases of parameter $\xi_1$ as well as for the standard $\Lambda$CDM model and the braneworld model at three selected values of redshift $z$.  }
	\label{fig:Table2} 
	\begin{tabular}{C{1.45cm} C{1.7cm} C{1.3cm} C{1.7cm} C{2.9cm} C{2cm}} \hline
		& & & & & \\[-2mm]
		\multirow{3}{*}{z} & \multicolumn{3}{c}{Ph-$\Lambda_{\rm s}$CDM model} & $\Lambda$CDM & \multirow{3}{*}{\makecell[c]{Braneworld\\model}} \\[2mm]
		&  $\xi_1=-1.5$ & $\xi_1=-1$ & $\xi_1=-0.5$ & $\xi=0,\xi_1=1$ & \\[2mm] \hline
		& & & & & \\[-1mm]
		2.33 & 218.72 & 221.49 & 225.77 & 236.63 & 231.14 \\[2mm]
		2.34 & 219.52 & 222.56 & 226.83 & 237.64 & 232.13 \\[2mm]
		2.40 & 225.17 & 229.03 & 233.20 & 243.73 & 238.09 \\[2mm] \hline
	\end{tabular}
\end{table}


\section{Conclusion}

In his talk \emph{“Future and Origin of our Universe: Modern View”}~\cite{bigrip1}, Alexei Starobinsky discussed the idea of a variable cosmological constant, for instance realized through a self-interaction potential $V(\phi)$ of a scalar field. The Ph-$\Lambda_{\rm s}$CDM~\cite{our} model provides a concrete realization of this concept: a phantom scalar field with a potential constructed to induce a rapid transition from an anti-de Sitter (AdS) phase to a de Sitter (dS) phase in the late Universe, thereby offering a physically well-defined embedding of the $\Lambda_{\rm s}$CDM framework~\cite{gDE,LsCDM1,LsCDM2,LsCDM3} — an approach that has shown promise in simultaneously addressing major cosmological tensions. Phantom fields, characterized by a negative kinetic term, violate the weak energy condition (WEC). In the same talk, Starobinsky discussed scenarios in which matter violates the WEC and emphasized that, if the equation-of-state parameter satisfies $w < -1$ and remains constant, the Universe will inevitably encounter a big rip singularity, where both the scale factor $a(t)$ and the energy density diverge within a finite cosmic time~\cite{bigrip2}. In our model, the phantom field indeed exhibits $w_\phi < -1$ after the transition. Nevertheless, the Universe asymptotically approaches a de Sitter phase, as in our model the EoS parameter satisfies $w_\phi \to -1$ in the limit $a \to \infty$, thereby evading the big rip singularity~\cite{bigrip4}. Moreover, since the total energy content includes cold dark matter and radiation, the total energy density $\varepsilon_{\rm tot}$, as well as the total inertial mass density $\varrho_{\rm tot} = \varepsilon_{\rm tot} + p_{\rm tot}$, remain positive throughout cosmic evolution. Hence, the WEC is not violated at the level of the total energy-momentum tensor, and the model remains physically viable.

Another interesting feature of our model is that the Hubble parameter $H(z)$ takes on smaller values than in the standard Planck-$\Lambda$CDM model prior to the transition. When combined with the CMB constraints given in Eqs.~\eqref{eq:Omega_M0},~\eqref{eq:Omega_r0}, and~\eqref{eq:D_M}, this feature enables the model to yield a Hubble constant $H_0$ that is in agreement with the SH0ES measurement. This outcome arises because the comoving angular diameter distance to the last scattering surface, $D_M(z_*)$, is kept fixed at the Planck-$\Lambda$CDM value. In our model, the Universe's evolution prior to recombination follows standard cosmology, ensuring that the sound horizon scale $r_*$ remains unchanged. As a result, the integral in Eq.~\eqref{eq:D_M} imposes a strict constraint on the expansion history: any suppression of $H(z)$ at redshifts before the transition must be compensated by an enhancement at later times. This built-in mechanism naturally allows for a higher value of $H_0$ without disrupting consistency with early-universe observables.

In Ref.~\cite{Starob}, Varun Sahni, Arman Shafieloo, and Alexei Starobinsky investigated the cosmic dynamics in a braneworld framework~\cite{Brane}, in which the cosmological constant is dynamically screened at early times. A key prediction of their model is that the Hubble parameter $H(z)$ remains consistently lower than in the Planck-$\Lambda$CDM model over the entire redshift interval from today ($z = 0$) to recombination. This includes the epoch around $z = 2.34$, where BAO measurements from the Ly-$\alpha$ forest are available. In the present work, we revisit this analysis by imposing the additional constraints given in Eqs.~\eqref{eq:Omega_M0}–\eqref{eq:D_M}, which were not considered in the original study. This refinement ensures that the braneworld model remains consistent with the full Planck-CMB dataset while simultaneously accommodating the SH0ES measurement of the Hubble constant. As shown in Fig.~\ref{fig:H_over_z}, these constraints cause the $H(z)$ curve to rise above that of $\Lambda$CDM at low redshifts ($z < 1$), enabling the model to reach $H_0 = 73.04\, \mathrm{km\,s^{-1}\,Mpc^{-1}}$, in line with local determinations.

One of the main findings of Ref.~\cite{Starob} was that the predicted Hubble parameter at redshift $z = 2.34$ in the braneworld model provides a better fit to the BAO measurement from Ly-$\alpha$ forest data~\cite{Delubac et al 2015} than the corresponding prediction from the Planck-$\Lambda$CDM model. In our paper, we set out to explore whether a similar conclusion can be drawn for our dynamical dark energy model and for the revised braneworld scenario under updated observational constraints. At the time of Ref.~\cite{Starob}, the measurement of $H(z = 2.34)$ from~\cite{Delubac et al 2015} was the only available BAO determination in that redshift regime. Since then, several new measurements have been reported in the range $z = 2.3$–2.4, as compiled in Table~\ref{tab:Table1}. These updated results, based on both SDSS and DESI data, exhibit some variation depending on the dataset and methodology employed. In Fig.~\ref{fig:H_over_z}, we have plotted these BAO-inferred values of $H(z)$ with corresponding error bars. A visual comparison shows that the SDSS-based measurements continue to favor expansion histories in which the Hubble parameter is suppressed at high redshift — such as those predicted by our Ph-$\Lambda_{\rm s}$CDM and braneworld models — whereas the most recent DESI measurement at $z = 2.33$ aligns more closely with the standard $\Lambda$CDM model. Thus, current data do not allow for a definitive conclusion regarding which model best matches the high-redshift BAO measurements.

In this brief paper, written in memory of Alexei Starobinsky, we revisited the braneworld model originally studied by Starobinsky and collaborators, which at the time was motivated by the apparent tension between Ly-$\alpha$ BAO data and $\Lambda$CDM. Although this anomaly appears less significant in light of more recent data, we find that the same model — once properly constrained — offers a natural resolution to the Hubble tension, a problem that gained prominence only after the original work was published. This capacity to anticipate the future relevance of ideas is a hallmark of scientific foresight — and a pattern often seen in the work of truly great minds.

\begin{acknowledgement}
This work was partially supported by the Center for Advanced Systems Understanding (CASUS) which is financed by Germany's Federal Ministry of Education and Research (BMBF) and by the Saxon state government out of the State budget approved by the Saxon State Parliament. A.Z. is also grateful to ICTP, Trieste Observatory and IFPU for their hospitality during the preparation of the final version of the paper.
\end{acknowledgement}

\end{document}